\documentclass[a4paper,11pt]{article}
\usepackage{authblk}
\usepackage{graphicx}
\usepackage{amsmath}
\usepackage{amsfonts}
\usepackage{amssymb}
\usepackage{amsmath,stackrel}

\begin{document}
\date{\today}

\title{~ \hfill {\small IMSc/2023/03} \\ [1.5cm]
Impact of errors in the magnetic field measurement on the
precision determination of neutrino oscillation parameters at the proposed ICAL detector at INO}

\author{Honey Khindri\thanks{Corresponding Author; honey@tifr.res.in}~}
\author{D. Indumathi\thanks{indu@imsc.res.in}}
\affil{Homi Bhabha National Institute,\\
Anu Shakti Nagar, Mumbai, India, and}

\affil{The Institute of Mathematical Sciences,\\
Taramani, Chennai, India}

\author{Lakshmi S. Mohan\thanks{Lakshmi.Mohan@ncbj.gov.pl}}

\affil{National Centre for Nuclear Research (NCBJ),\\
Warsaw, Poland}

\maketitle

\begin{abstract}
The magnetised iron calorimeter (ICAL) detector proposed at the
India-based Neutrino Observatory will be a 51 kton detector made up of
151 layers of 56 mm thick soft iron with 40 mm air gap in between
where the RPCs, the active detectors, will be placed. The main goal
of ICAL is to make precision measurements of the neutrino oscillation
parameters using the atmospheric neutrinos as source. The charged current
interactions of the atmospheric muon neutrinos and anti-neutrinos in the
detector produce charged muons. The magnetic field, with a maximum value
of $\sim$ 1.5 T in the central region of ICAL, is a critical component
since it will be used to distinguish the charges and determine the
momentum and direction of these muons. It is difficult to measure the
magnetic field inside the iron. The existing methods can only estimate the
internal field and hence will be prone to error. This paper presents the
first simulations study of the effect of errors in the measurement of the
magnetic field in ICAL on its physics potential, especially the neutrino
mass ordering and precision measurement of oscillation parameters in the
2--3 sector. The study is a GEANT4-based analysis, using measurements of
the magnetic field at the prototype ICAL detector. We find that there
is only a small effect on the determination of the mass ordering. While
local fluctuations in the magnetic field measurement are well-tolerated,
calibration errors must remain well within 5\% to retain good precision
determination of the parameters $\sin^2\theta_{23}$ and $\Delta m^2_{32}$.

\end{abstract}



\maketitle

\section{Introduction}

The proposed magnetised Iron Calorimeter (ICAL) detector at the
India-based Neutrino Observatory (INO) \cite{ICAL:2015stm} is primarily
designed as an atmospheric neutrino detector. Its main goal is to detect
muons produced in the charged current (CC) interactions of atmospheric
muon neutrinos (and anti-neutrinos) with the detector, via
\begin{equation}
\nu_\mu N \to \mu^- X~; \hspace{1cm}
\overline{\nu}_\mu N \to \mu^+ X~.
\end{equation}
The magnetic field will determine the sign of the charge of the muons
(through the bending of the charged particle tracks in the detector) and
hence be able to distinguish neutrino- and anti-neutrino-initiated
events. This key feature will enable a clean determination of the as-yet
unknown sign of the neutrino mass ordering and hence the neutrino mass
hierarchy. A precise determination of the magnetic field map over the
entire ICAL detector is therefore a crucial input in this determination.
The ICAL detector will also make precision measurements of the 2--3
neutrino oscillation parameters such as $\sin^2\theta_{23}$ and $\Delta
m^2_{32}$; both these also depend on correctly determining the muon
momenta using the knowledge of the magnetic field map. In this paper,
we present for the first time, a detailed simulations study of
the impact of measurement errors in the magnetic field on the physics
potential of the ICAL detector.

The paper is organised as follows: we present a short summary of the
current status of neutrino oscillation physics in Section 2. In Section
3, we present some highlights of the proposed ICAL detector. We also
present a summary of the magnetic field measurements made in the prototype
mini-ICAL detector, an 85 ton scale model, that has been functioning for
the last few years. We will use these results, especially, the measured
errors in both calibration and measurement of the magnetic field,
in Section 5, to analyse the impact of these errors on the precision
measurement of the neutrino oscillation parameters. Before doing this,
we first present in Section 4 a detailed simulations study of the
effect of errors---in calibration and in measurement---of the magnetic
field map on the reconstruction of the muon momenta and compare it with
older results where the magnetic field was assumed to be well-defined
\cite{Chatterjee:2014vta}. We use these results to parametrise the changes
in the reconstruction values (central value, spread, etc), with respect
to fixed changes in the magnetic field. We present our results in Section
6 and We conclude with a discussion of the results in Section 7.

\section{Neutrino oscillation: summary and status}

In 1968 Pontecorvo \cite{pontecorvo1968neutrino} proposed the quantum
mechanical phenomenon of neutrino oscilaations in analogy with K$^0$ and
$\overline {K^0}$ oscillations. In 1962 Maki, Nakagawa and Sakata
\cite{10.1143/PTP.28.870} for the first time constructed a model with
mixing of different neutrino flavors. Currently, the three neutrino
flavours, $\nu_e, \nu_\mu, \nu_\tau$, can be expressed in terms of the
mass eigen states, $\nu_1, \nu_2, \nu_3$, using three mixing angles
$\theta_{12}$, $\theta_{23}$ and the across-generation mixing angle,
$\theta_{13}$, as well as a CP-violating phase for Dirac-type neutrinos,
$\delta_{CP}$.  Non-zero and different neutrino mass-squared differences,
$\Delta m_{ij}^2 \equiv m_i^2 - m_j^2$, with $i,j=1,2,3$ along with mixing
lead to neutrino oscillations which have since been observed in solar,
atmospheric, reactor, and accelerator neutrinos.

\subsection{Neutrino experiments}

The first solar neutrino experiment were detected by Ray Davis and his
collaborators in the Homestake experiment using Chlorine as the active
detector material. Neutrino oscillation could account for the fact that
the measured rate of charged current interactions of electron-type
neutrinos was one third of the rate predicted by the solar standard
model. This was followed by Kamiokande and Super Kamiokande experiments
with water Cerenkov detectors in Japan, and lastly by the Sudbury
Neutrino Observatory in Canada which observed the solar neutrinos in
both the charged and neutral current (NC) channels and confirmed the
phenomenon of neutrino oscillations and non-zero neutrino masses in the
solar (primarily 1--2) sector. The anomaly between the atmospheric
electron and muon neutrino measurements of the Super Kamiokande
experiment could also be resolved within the neutrino oscillation
paradigm.

Subsequently, T2K long baseline neutrino experiment, the LSND
experiment, the Daya Bay reactor (anti-)neutrino experiment, the
MiniBoone experiment, the MINOS long baseline experiment and the Ice
Cube experiment at the South pole have all confirmed neutrino
oscillations in different flavours and sectors. While the solar and
reactor neutrino sector mainly determines the parameters in the 1--2
sector, atmospheric and accelerator experiments have been used to pin
down the parameters of the 2--3 sector. In particular, the reactor
experiments have played a crucial role in determining the non-zero
though small value of $\theta_{13}$. While it has been established that
$\Delta m^2_{21} > 0$, the mass ordering in the 2--3
sector, viz., the sign of the mass squared difference $\Delta m^2_{32}$
(or equivalently, of $\Delta m^2_{31}$), is currently not known. In
addition, the value of the CP phase as well as the octant of the mixing
angle $\theta_{23}$ is not yet established. The up-coming DUNE 
\cite{DUNE:2020jqi} and JUNO \cite{JUNO:2022mxj}
experiments will also probe these parameters. The current status of
neutrino oscillation mixing parameters can be found in
Refs.~\cite{ParticleDataGroup:2020ssz,Esteban:2020cvm}; the best-fit
values and their $3\sigma$ ranges (which we use in this
analysis\footnote{The values have been updated since then; however, for
the convenience of comparing with our older results where we have
assumed no errors on the magnetic field map, we are using the values
given in this reference, Ref.~\cite{Esteban:2020cvm}.}) are shown in
Table \ref{tab:osc}.

For convenience, we define \cite{Senthil:2022tmj}
\begin{equation}
\Delta m^{2} \equiv m_{3}^{2}-\left(\frac{m_{2}^{2}+m_{1}^{2}}{2}\right)~.
\label{eq:dm2}
\end{equation}
Note that $\Delta m^2$ flips sign {\em without changing its magnitude}
when the hierarchy/ordering changes and hence is a convenient parameter
compared to $\Delta m_{31}^2$ or $\Delta m_{32}^2$, which change in both
sign and magnitude depending on the mass ordering. We shall use $\Delta
m^2$ throughout in the analysis. Depending on the mass ordering, and using
the value of $\Delta m_{21}^2$ given in Table \ref{tab:osc}, the values
of $\Delta m_{31}^2$ and $\Delta m_{32}^2$ can be found from $\Delta
m^2$. We shall assume the normal ordering throughout this analysis,
unless otherwise specified.

\begin{table}[htp]
\centering
\vspace{0.1cm}
\begin{tabular}{| c | c | c |}
\hline
 Parameter & Central values & $3 \sigma $ Range \\ \hline
$\sin^2 \theta_{12}$ & $0.304$ & fixed \\
$\sin^2 \theta_{13}$ &$0.0222$ & $0.0203 \leftrightarrow 0.0241$\\
$\sin^2 \theta_{23}$ & $0.5$ & $0.381 \leftrightarrow 0.615$ \\
$\Delta m^2_{21}$ $(\times 10^{-5}$ eV$^2$) & $7.42$ & fixed \\
$\vert \Delta m^2 \vert$ $(\times 10^{-3}$ eV$^2$) & $2.47$ & $2.395 \leftrightarrow
2.564$  \\
$\delta_{CP} (^\circ)$ & $0.0$ & fixed \\ \hline 
\end{tabular}
\caption{The $3\sigma$ ranges of neutrino oscillation parameters --- mixing
angles and mass squared differences --- and central values used in the
present work \cite{ParticleDataGroup:2020ssz,Esteban:2020cvm}; $\Delta
m^2$ is defined in Eq.~\ref{eq:dm2}.}
\label{tab:osc}
\end{table}

\section{Highlights of the ICAL detector}

There are two sources of muon neutrinos (and anti-neutrinos) from
Earth's atmosphere. One are those muon neutrinos produced from decays of
secondary pions via $\pi \to \mu \nu_\mu$ with the subsequent decay of
the muons into additional muon neutrinos via $\mu \to \nu_\mu e \nu_e$.
Due to the presence of neutrino oscillations, it is also possible to
detect muon neutrinos, which have been produced through oscillation of
electron neutrinos on their way to the detector. Typically, neutrinos
arriving from below (so-called up-going neutrinos, produced in the
atmosphere on the other side of the Earth) are more likely to exhibit
oscillations due to the relevant GeV-scale energies and path-lengths
involved. Hence there will be two contributions to the detected muons
arising from CC interactions in ICAL: those produced by muon neutrinos
that have survived during their journey, involving the neutrino survival
probability $P_{\mu\mu}$ and those produced by electron
neutrinos that have oscillated into muon neutrinos, involving the
oscillation probability $P_{e\mu}$. These probabilities depend on both
the neutrino energy and path-length travelled; depending on the neutrino
mass ordering, these show MSW enhancement 
\cite{Wolfenstein:1977ue, Mikheev:1986wj} in the few-GeV energy range
for neutrinos passing through Earth matter with
path lengths such that the zenith angle is less than $\theta <45^\circ$
(with $\cos\theta = 1$ for vertically upward neutrinos). Specifically,
the resonance will be visible in the neutrino sector for normal ordering
with $\Delta m^2 >0$ and in the anti-neutrino sector for inverted
ordering with $\Delta m^2 <0$.

\subsection{The ICAL detector}

The proposed ICAL (Iron Calorimeter) detector is a 51 kton magnetised
detector to be located at the India-based Neutrino Observatory (INO)
with a rock cover of at least 1 km in all directions. It will consist
of three modules of size 16  m $\times$ 16 m $\times$ 14.5 m in ($x$,
$y$, $z$ dimensions) consisting of 150 layers of resistive plate chamber
(RPC) which will act as an active detector to detect muons and 151 layers
of 56 mm thick iron plates which will act as interaction material for
neutrinos. There will be a 40 mm gap between each two iron layers to
place the RPC in between. There will be three sets of current carrying
copper coils which will be energised to produce the required magnetic
field in the detector.

\begin{figure}[hbp]
\centering
\includegraphics[width=0.65\textwidth]{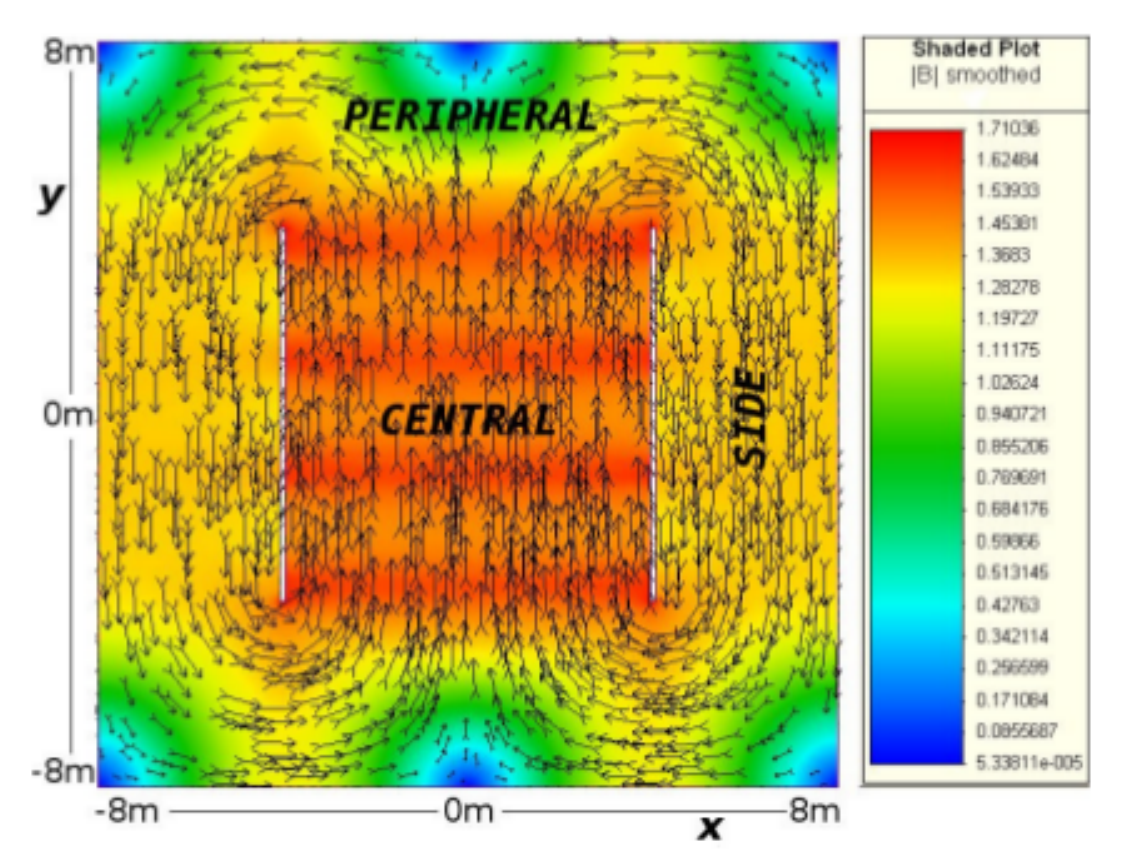}
\caption{Field map of simulated magnetic field for ICAL detector from
Ref.~\cite{shibu}; shown
is the field in the $x$--$y$ plane in the centre of a single iron layer.
The field is uniform in the $z$ direction over the thickness of the
iron.}
\label{fig:field_map}
\end{figure}

ICAL is designed to detect muons of energy range from 1--25 GeV generated
from the CC interactions of the atmospheric $\nu_{\mu}$
and $\overline{\nu}_{\mu}$ neutrinos. One of the special features of ICAL is
that it is sensitive to the charge of muons because of the presence of
magnetic field of B$_{max}$ $\sim$ 1.5 Tesla. While the field is mainly
in the $y$ direction in the central region (see
Fig.~\ref{fig:field_map}) and in the side region, although smaller
by about 15\%, it varies in both magnitude and direction in the
peripheral region. The map has been generated \cite{shibu} using the MAGNET6
\cite{magnet6} software and has been extensively used in all simulations
analyses of ICAL \cite{ICAL:2015stm}.

Each RPC has pick-up strips along the $x$- and $y$-directions respectively
above and below, so that whenever a charged particle passes through
it, signals will be collected as ``hits" for that layer independently
in both the $x$ and $y$ directions. As the muon bends in the magnetic
field depending on the sign of its charge, the bending of the track along
with information on the magnetic field in the region is used by a Kalman
Filter program to reconstruct the muon momentum (magnitude and direction)
as well as the sign of its charge, depending on whether it is up-coming
or down-going; this latter is determined based on the timing information
which is available to an accuracy of about 1 ns. We will present results
on the muon reconstruction efficiencies and energy resolutions in a
later section. We will first present details of the measurement of the
magnetic field at the prototype mini-ICAL detector and how we use it as
an input in the current analysis.

\subsection{Experimental inputs from mini-ICAL}

\begin{figure}[!tbp]
  \centering
   \includegraphics[width=0.65\textwidth]{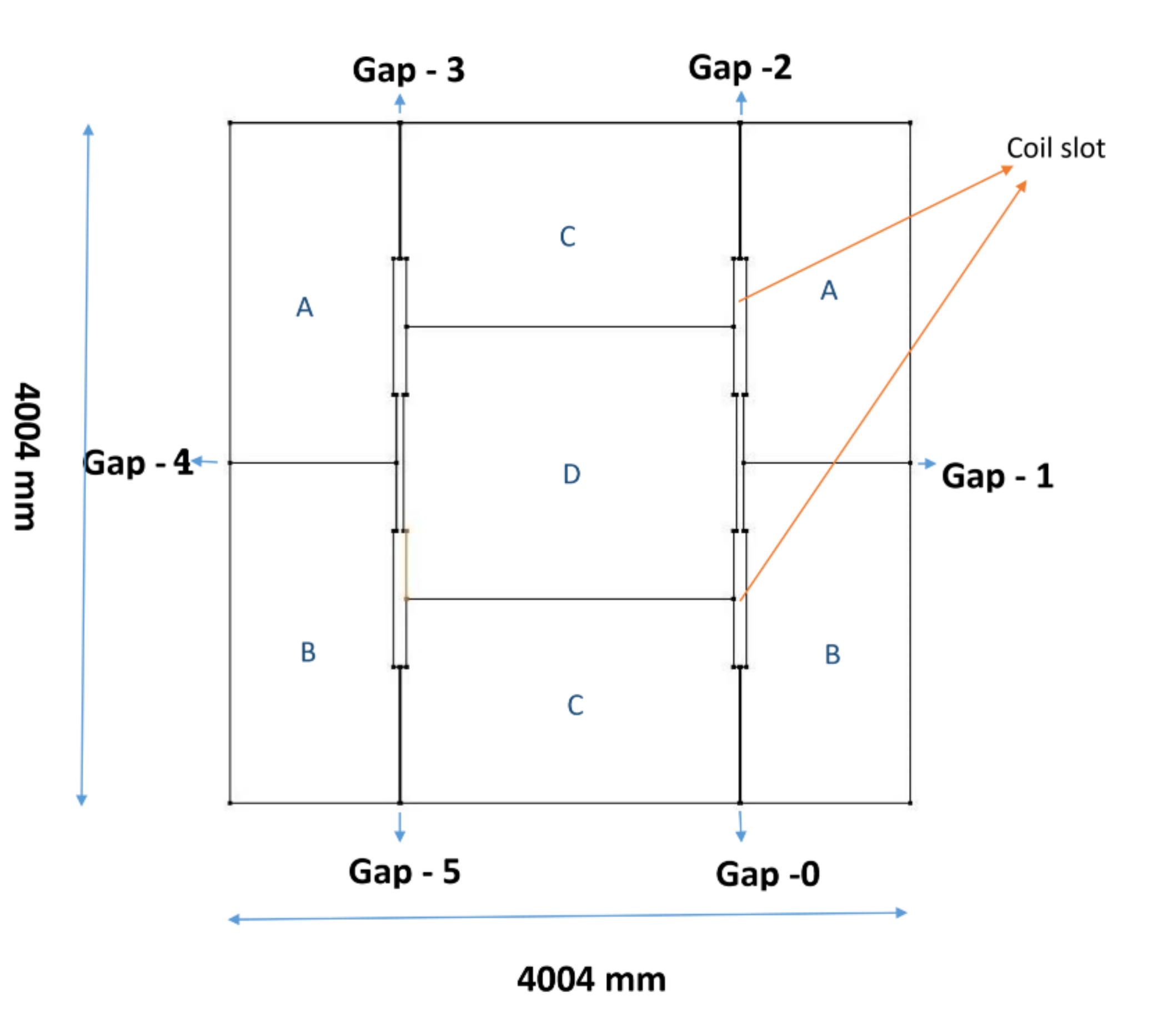}
    \caption{Schematic of top view of {\sf mini-ICAL} showing the gaps
    between plates in a layer.}
    \label{fig:mini_ical_up}
\end{figure}

The mini-ICAL is a fully functional prototype detector working at the
transit campus of INO near Madurai in Tamil Nadu, India. It was built to
study the challenges and difficulties that could be faced while installing
the main ICAL detector. It is an 85 ton detector which consists of 11
layers of 56 mm thick iron and 10 layers of RPCs which are sandwiched
between the iron layers as active detector elements. Each iron layer is
made up of 7 plates with an overall dimension of $4 \times 4$ m$^2$; see
Fig.~\ref{fig:mini_ical_up}. The two sets of copper coils each having
18 turns are used to magnetize mini-ICAL by flowing current through
the coils.

\begin{figure}[h]
 \centering
\includegraphics[width=0.49\textwidth]{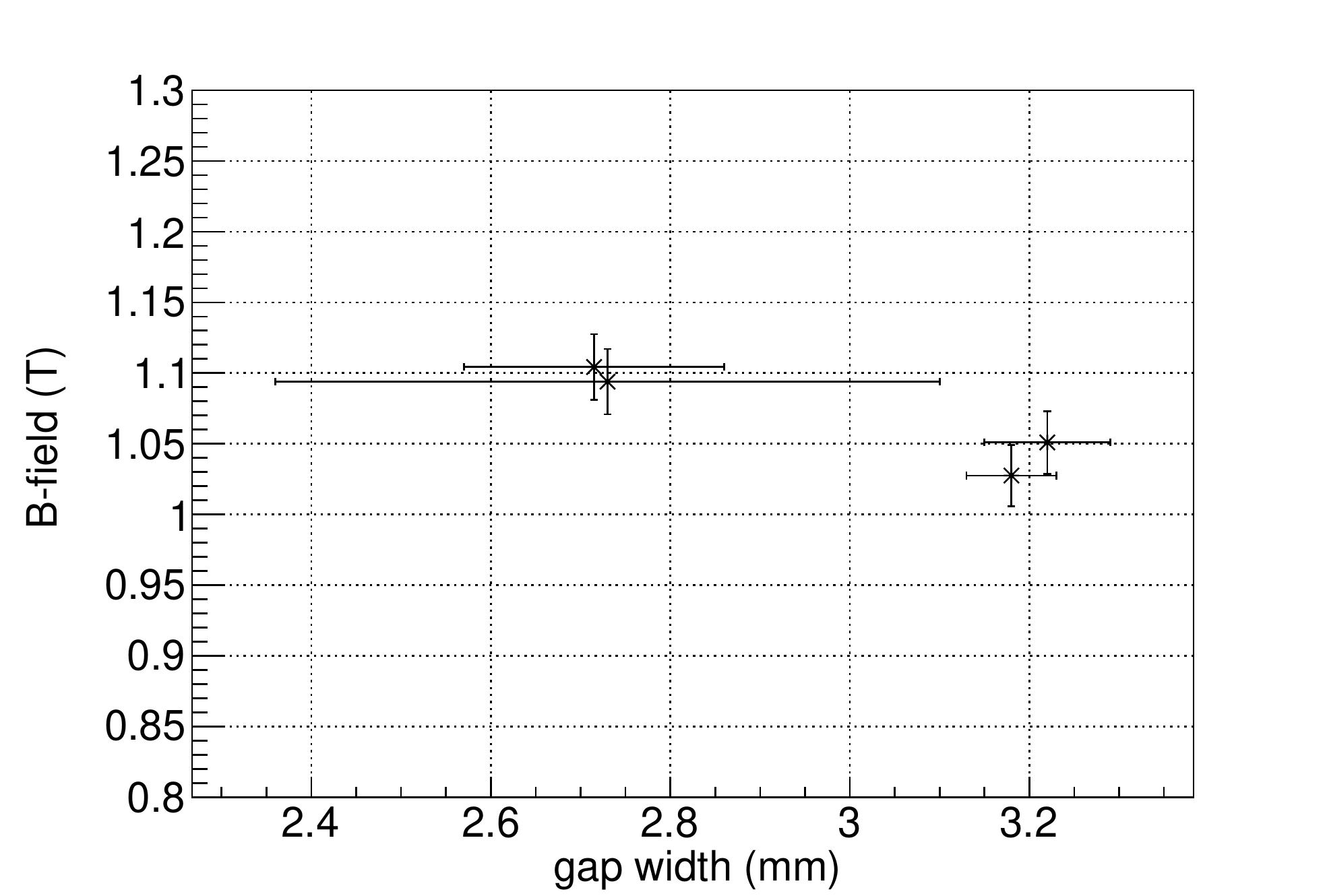}
\includegraphics[width=0.49\textwidth]{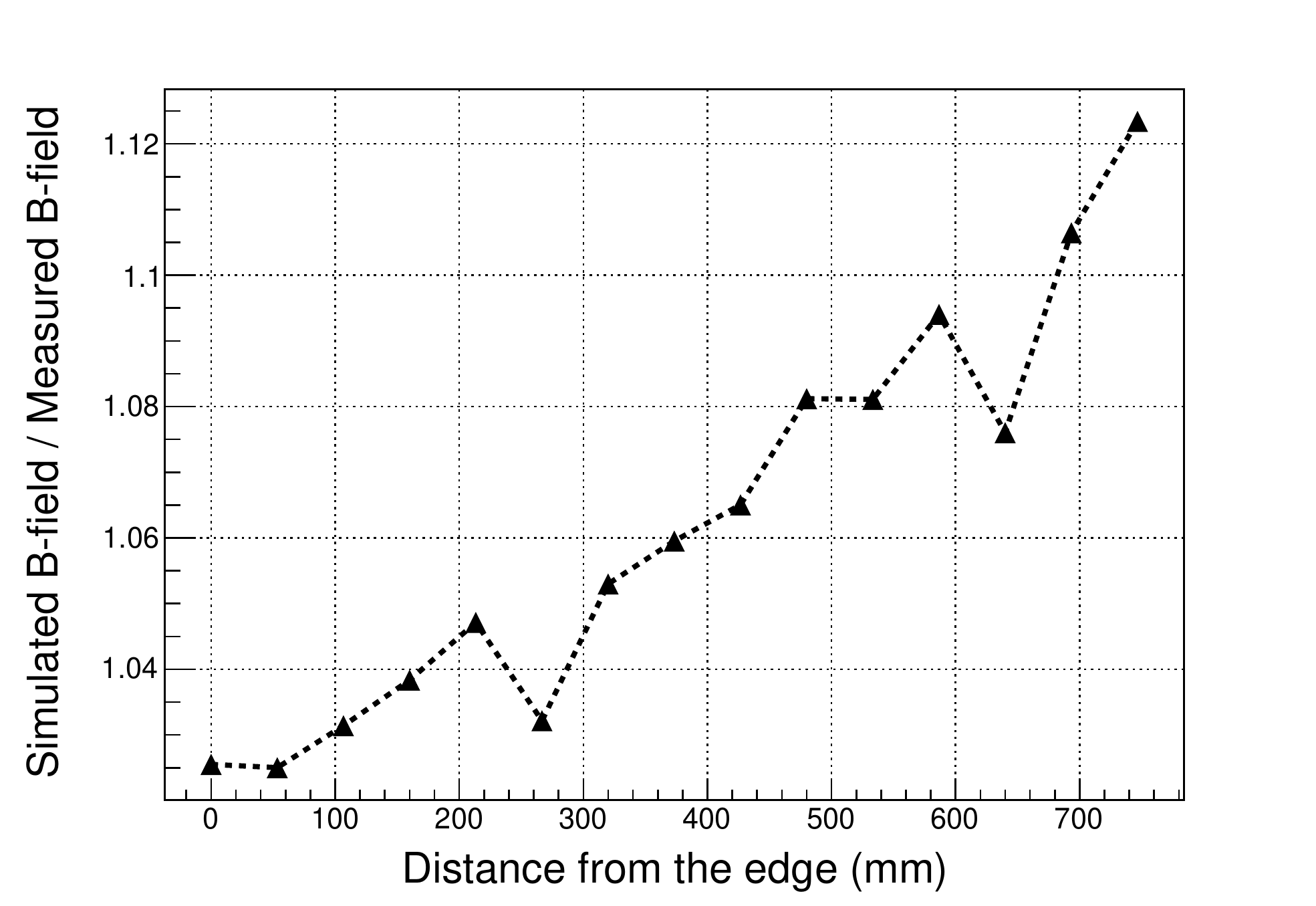}
\caption{Left: Variation of magnetic field as a function of the gap
width in mini-ICAL, from Ref.~ \cite{Khindri:2022elz}. Right: Ratio of
simulated
\cite{Honey_sim} to measured magnetic field in gap 0 of mini-ICAL as a
function of the distance from the edge of the detector. The right-most
values are close to the coil gap.}
\label{fig:mini_ical_field}
\end{figure}

The mini-ICAL detector has been constructed with 3 mm gaps built-in
on purpose between the different iron plates in layers 1, 6, and 11
(top, middle and bottom); see
Fig.~\ref{fig:mini_ical_up}, in order to enable the measurement of
the magnetic field in the gaps using a Hall probe. A detailed study
of the magnetic field all along the gaps from a location near the coil
slots to the outer edges of the detector, and in the air just outside,
has been made \cite{Khindri:2022elz}. Separately, a detailed simulations
study of the magnetic field in mini-ICAL \cite{Honey_sim} as a function
of the coil current and the gap widths, has been made using the MAGNET6
software \cite{magnet6}. As expected, the magnetic field was more or
less uniform and largest in magnitude in the {\em central region},
with $\vert x \vert, \vert y \vert \le 1$ m between the coil slots.
(Note that the origin is defined to be the centre of the detector). The
field is somewhat smaller and in the opposite direction in the {\em side
region} outside the slots, $\vert x \vert \ge 1$ m; $\vert y \vert \le
1$ m, while the field is changing in both magnitude and direction in
the {\em peripheral region} with $\vert y \vert \ge 1$m. In particular,
in the peripheral region, the field is largest near the coil slot and
decreases toward the outside. Also, as expected, the field is smaller
when the gap width is larger; see Fig.~\ref{fig:mini_ical_field}. Hence
the field in the mini-ICAL is similar to that expected from simulations
on the main ICAL detector. In addition, it was found
\cite{Khindri:2022elz}.
that the magnetic field in mini-ICAL can be measured to within an accuracy
of 3\% \cite{Khindri:2022elz}.

A comparison of the measured and simulated field at mini-ICAL indicated
that it may be possible to get agreement between measured and simulated
field values to within about 10\%. When the difference between the
measured and simulated gap widths was taken into account, the agreement
improves to within 5\%. In this paper, we use the results of this study
and assume a similar result will hold for the main ICAL detector; i.e.,
we use the magnetic field map for the entire ICAL detector as simulated
by the MAGNET6 software. Assuming that errors on this do not exceed about
5\%, we present a preliminary and first study of the effect of errors
in the measurement of the magnetic fields on the physics goals of ICAL.

We use these values in our simulations study of muons in the main ICAL
detector, with true and modified magnetic fields.

\section{Simulations Study of Muons}

The GEANT4 \cite{geant} code is used to generate the ICAL geometry which comprises
three modules of 151 layers of 56 mm iron, separated by a 40 mm gap
in which the active detector elements, the RPCs, are inserted. The
magnetic field map shown for a single iron layer in Fig.~\ref{fig:field_map}
is generated using MAGNET 6.0 software \cite{ICAL:2015stm}. The calibration
of muon energy/momentum is done analogous to the approach discussed in
detail in Refs.~\cite{Chatterjee:2014vta,Kanishka:2015qsa}.
Here, 10,000 muons of fixed
energy and direction $(p_\mu,cos\theta)$ are generated at random over
the entire detector and propagated in the ICAL detector using the
magnetic field map shown in Fig.~\ref{fig:field_map}; henceforth this
is called the ``true" magnetic field, $B(x,y,z)$. The charged particles,
on passing through the RPCs, trigger a discharge, which is acquired as a
``hit" in the detector. The collection of hits through different layers
forms a track. The magnetic field bends the muons into the observed
muon track. The hit pattern is studied and the reconstruction of the
various muon properties is done using a Kalman filter algorithm which
requires hits in at least 5 consecutive layers. The algorithm identifies
the hits which are a part of the muon track (non-trivial in the case of
a genuine neutrino charged-current (CC) event with associated hadron
hits), and returns the direction $(\cos\theta, \phi)$, the magnitude of
the momentum and the sign of the charge of the muon. Various selection
criteria are used \cite{Chatterjee:2014vta,Kanishka:2015qsa}
to improve the fit quality. The
histogram of reconstructed momenta is fitted to a gaussian which returns
the mean reconstructed momentum/energy ($E_{reco} (p_\mu,\cos\theta)$)
and width ($\sigma (p_\mu, \cos\theta)$) as a function of the true input
values. The number of reconstructed events to the total number is the
reconstruction efficiency, $\epsilon_{reco}$, also a function of the input
energy/momentum and direction, and the relative number of correctly charge
identified tracks to the number reconstructed is the relative charge
identification efficiency, $\epsilon_{cid}$. This analysis was performed
for sets of muons with energies 1--25 GeV in steps of 1 GeV, and with
zenith angle $\cos\theta = 0.3$--0.9 in steps of 0.1. Details of the
quality of these fits, etc., can be found in the references given earlier.

\subsection{Parametrisation with modified magnetic fields}

We wish to study the response of the detector when the measured magnetic
field is different from the true one. In order to be able to parametrise
the effect of such errors, and to quantify them, we use the following
approach.

\begin{figure}[hbp]
\centering
\includegraphics[width=0.65\textwidth]{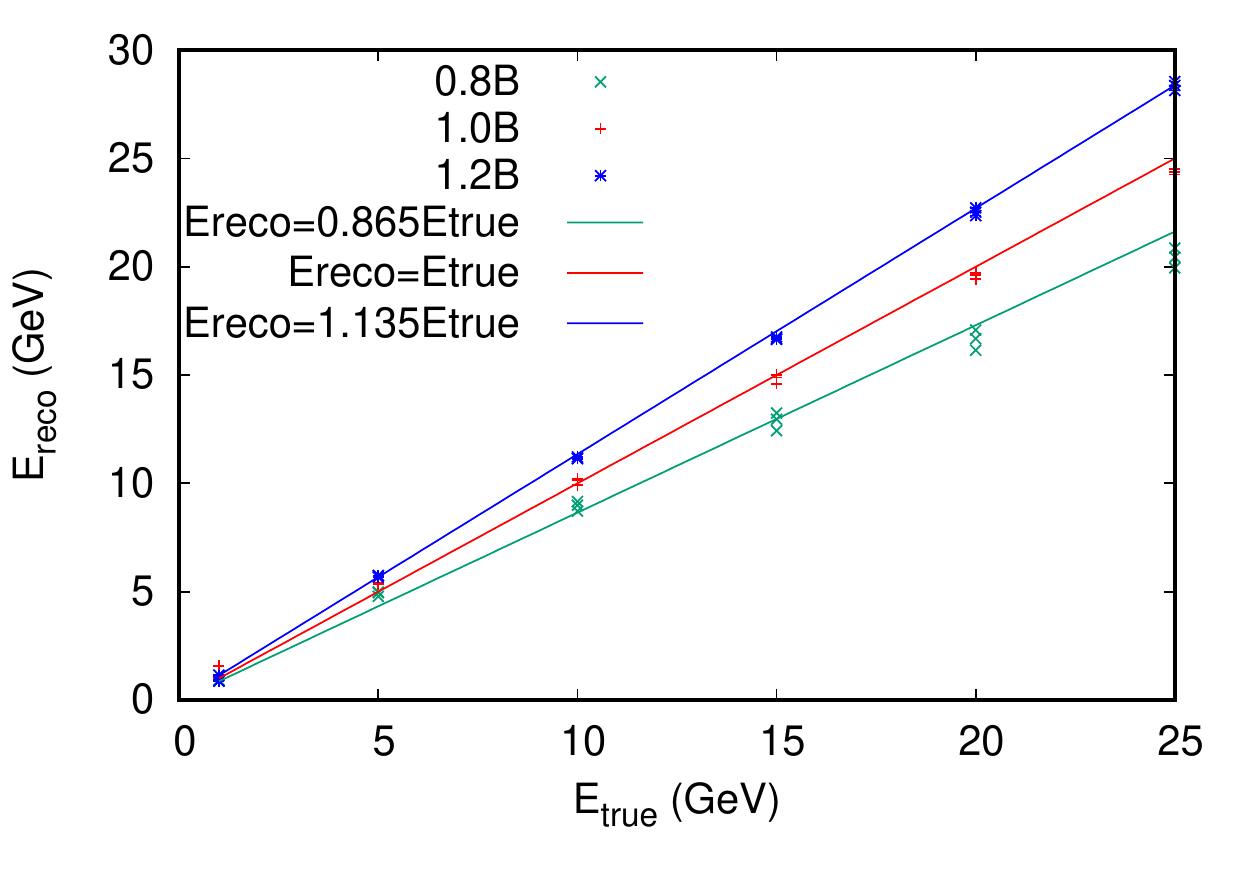}
\caption{Reconstructed muon energy E$_{reco}$ vs E$_{true}$ for
cos$\theta$ = 0.4, 0.6 and 0.8 using a magnetic field which is 0.8,
1.0 and 1.2 times the true magnetic field map.}
    \label{fig:Ereco}
 \end{figure}

The same muon tracks are now fitted and reconstructed with a computed or
simulated magnetic field map which is different from the actual one. For
simplicity, 6 scenarios, when the fitted magnetic field is systematically
different from the true value by a constant factor, $B_{reco} = fB$,
are considered, viz., the fields are taken to be $B_{meas} = 0.7B, 0.8B,
0.95B, 1.05B, 1.2B, 1.30B$, so $f=0.7,0.8,0.95,1.05,1.2,1.3$, that is,
magnetic fields which are 30\%, 20$\%$, or 5$\%$ smaller or larger
than the true magnetic field as given by the magnetic field map. The
reconstructed energy $E_{reco}$, the energy resolution $\sigma$, and
the reconstruction and charge identification efficiencies are calculated
in each case. Fig.~\ref{fig:Ereco} shows the reconstructed muon energy
as a function of the true value over the relevant range for
atmospheric neutrinos from 1--25 GeV. In the interests of clarity, only
every 5th energy value is plotted for angles $\cos\theta = 0.4, 0.6,
0.8$ (closely overlapping points with the smallest value corresponding
to the smallest $\cos\theta$). It can be seen that the reconstructed
energy $E_{reco}^{fB}$ increases (decreases) for a given true value
$E_{true}$ as $f$ increases (decreases). The results for $f=0.8, 1.2$ only
are plotted in comparison to the true value ($f=1$), again for clarity.

\begin{figure}[hbp]
\centering
\includegraphics[width=0.65\textwidth]{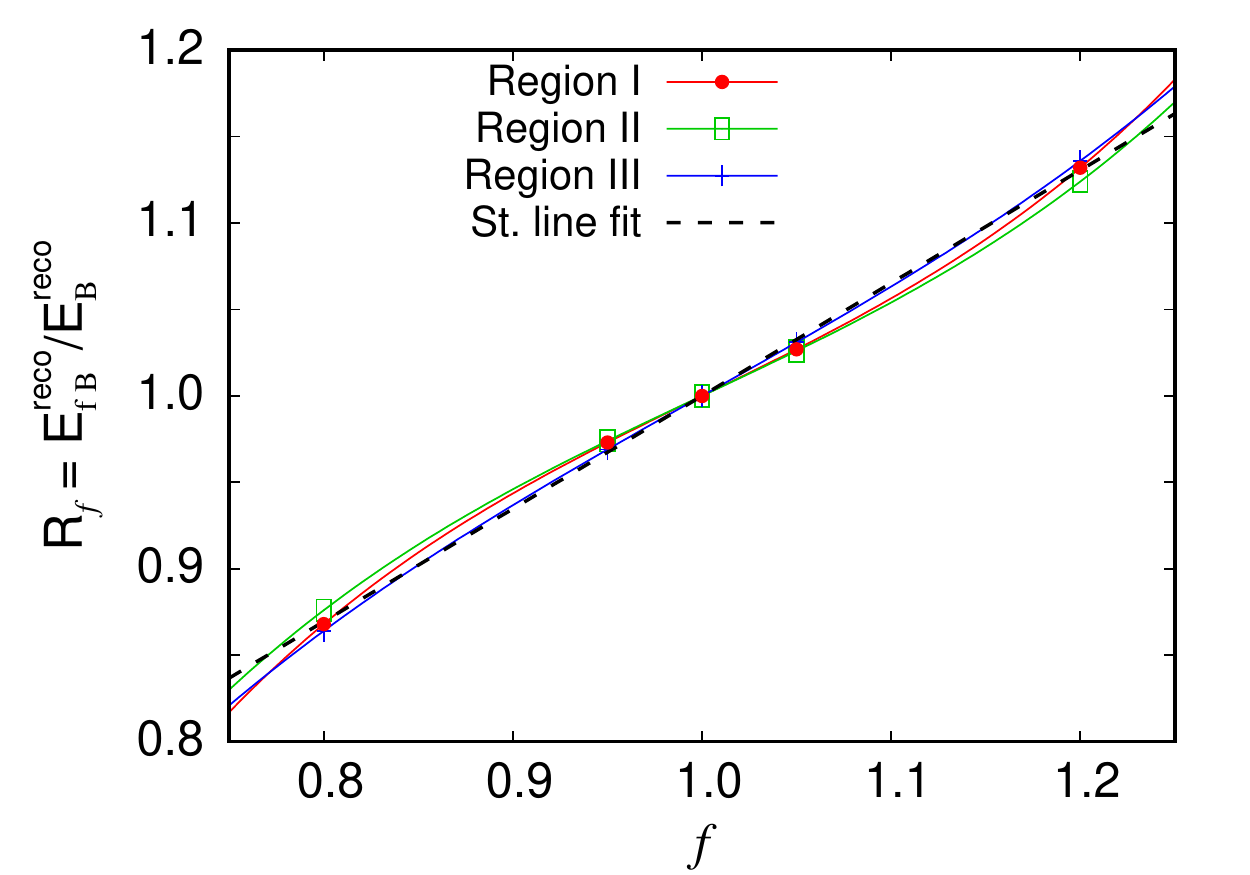}
\vspace{-4mm}
\caption{The ratio $R_f$ of the reconstructed muon energies when the
modified and true magnetic field maps respectively are used, as a
function of the modification parameter, $f$. The results are similar for
the three different regions, central (I), side (II), and peripheral
(III), as explained in the text. The ratio is almost linear for smaller
values of $\vert 1-f \vert$.}
    \label{fig:par}
 \end{figure}

It turns out that E$^{fB}_{reco}$ is given to a very good accuracy by a
constant scale parameter compared to E$^{B}_{reco}$ for the original map,
where $f$ = 0.8, 0.95, 1.05, 1.2, etc. The actual reconstructed values (for
$\cos\theta = 0.8$) are shown along with this fit in Fig.~\ref{fig:Ereco}
as a function of E$_{true}$. For larger deviation from the true magnetic
field ($\sim$20 \%), it can be seen from Fig.~\ref{fig:par} that there
is a non-linear variation of the reconstructed energy; the results fit
marginally better to a quadratic, but for smaller deviations (upto 5\%)
the fit is almost linear for all the three mentioned regions (I, II and
III), and therefore the change can be parametrised by fitting with a
straight line. This is shown in Table~\ref{tab:E_reco}.

\begin{table}[htp]
\centering 
\begin{tabular}{|c|c|c|} \hline  
B-field& f & $E_{reco}$ vs $E_{true}$  \\ \hline
$B = B_{true}$    & 1.00 & $E_{reco} \sim E_{true}$    \\ 
$B = 0.8B_{true}$ & 0.80 & $E_{reco} = 0.865 E_{true}$ \\
$B = 0.95B_{true}$& 0.98 & $E_{reco} = 0.965 E_{true}$ \\
$B = 1.05B_{true}$& 1.02 & $E_{reco} = 1.035 E_{true}$ \\
$B = 1.2B_{true}$ & 1.20 & $E_{reco} = 1.135 E_{true}$ \\ \hline
\end{tabular}
\caption{Change in the energy reconstruction (E$_{reco}$) of muon w.r.t.
change in the magnetic field by a constant factor $f$ across the entire
map.}
\label{tab:E_reco}
\end{table}

In addition the width and hence the energy resolution for muons varies
as $\sigma (f B) = f \sigma(B)$. The angular resolution, the
reconstruction and charge identification efficiencies are practically
independent of the choice of magnetic field map, except at very low
energies, $E_\mu \le 2$ GeV. In this first, rather simplistic analysis,
we ignore these outliers and take these values to be the same as with
the true magnetic field. Obviously, the hadron energy resolution (fitted
to the number of hadron hits; see Ref.~\cite{Devi:2018ltf}
for details) is also independent of the choice of the magnetic field map.

With this analysis, we are now ready to analyse the impact of magnetic
field measurement errors on the physics reach of ICAL from charged
current (CC) interactions of atmospheric neutrinos in the detector.
Throughout we use the baseline results from Ref.~\cite{Mohan_2017,Senthil:2022tmj} for comparison.

\section{Simulations study of impact of errors in magnetic field
measurement at ICAL}

\subsection{Events generation}

Both $\nu_e$ and $\nu_\mu$ types of neutrinos (and their anti-particles)
are present in atmospheric neutrino fluxes (the direct tau-neutrino
contribution is negligible at these energies and are only produced
through oscillation). Charged current (CC) atmospheric neutrino (and
anti-neutrino) events were simulated for an exposure of 1000 years at
the ICAL detector using the NUANCE neutrino generator \cite{Casper:2002sd}
and Honda 3-D fluxes \cite{Honda:2004yz}. The ``data" is generated using the
current central best fit values of various neutrino oscillation parameters
as listed in Table~\ref{tab:osc} which are considered to be the ``true"
values. Note that the analysis is completely insensitiveto the CP phase
$\delta_{CP}$ and we have assumed maximum mixing in the 2--3 sector,
viz., $\theta_{23} = 45^\circ; \sin^2\theta_{23} = 0.5$ throughout.
In addition, normal ordering is assumed unless otherwise specified.

Since the ``theory" events are to be generated with different values of
these oscillation parameters, the NUANCE events were generated without
oscillations, which were applied event-by-event later. The rate of
observed charged muons of either type in the detector in terms of the
true final state muon energy and direction is given by
\begin{eqnarray}
\frac{{\rm d}^2 N_{\mu}}{{\rm d}E_{\mu} {\rm d} \cos \theta_{\mu}} & = &
	T \times N_{D} \times \int {\rm d}E_{\nu} {\rm d}\cos\theta_\nu
	{\rm d}\phi_\nu \left[P_{\mu\mu} \frac{{\rm d}^3 \Phi_{\mu}}
	{{\rm d}E_{\nu} {\rm d}\cos\theta_{\nu} d\phi_{\nu}} + \right.
	\nonumber \\
	 & & \left. P_{e\mu} \frac{{\rm d}^3\Phi_{e}}{{\rm d}E_{\nu}
	{\rm d}\cos\theta_{\nu} {\rm d}\phi_{\nu}}
	\right] \times \frac{{\rm d}^2\sigma^{CC}_{\mu}}{{\rm d}E_\mu
	{\rm d}\cos\theta_\mu}~,
\end{eqnarray}
where $T$ is the exposure time in seconds, $N_D$ is the number of target
nuclei, $\Phi_{e,\mu}$ are the electron and muon type atmospheric
neutrino fluxes, and $\sigma^{CC}_{\mu}$ is the charged current cross
section for the muon neutrino interaction in the detector. Here
$P_{\mu\mu} (E_\nu,\cos\theta_\nu)$ is the muon neutrino survival
probability and $P_{e\mu} (E_\nu,\cos\theta_\nu)$ is the oscillation
probability of electron neutrino into muon neutrino. For anti-neutrinos,
the corresponding anti-neutrino fluxes, cross sections and probabilities
are used. It is seen that both $\nu_\mu$ and $\nu_e$ fluxes contribute
(the former through the survival probability $P_{\mu\mu}$ and the latter
through the oscillation probability $P_{e\mu}$). Hence events using both
sets of initial fluxes were generated.

Symbolically, the true number of oscillated events can be expressed as
\begin{eqnarray}
N_{\mu^-} & = & N_{\mu^-}^{0}\times P_{\mu\mu} +
N_{e^-}^{0}\times P_{e\mu}~, \\ \nonumber
N_{\mu^+} & = & N_{\mu^+}^{0}\times \overline{P}_{\mu\mu} +
N_{e^+}^{0}\times \overline{P}_{e\mu}~.
\label{eq:events}
\end{eqnarray}
That is, the events N$^{0}_{\mu\pm}$ were generated by using the
$\nu_{\mu}$ fluxes and the events N$^{0}_{e\pm}$ were generated by
swapping the $\nu_{\mu}$ and $\nu_e$ fluxes in the generator, retaining
the same cross sections.

The above equations, Eq.~\ref{eq:events}, are just representative in
order to understand the different contributions. Actually, the events are
generated by NUANCE, so that $N^0_{\mu\pm}$ are the un-oscillated muon
events and $N^0_{e\pm}$ are swapped muon events generated by NUANCE,
and $N_{\mu^\pm}$ are generated from the $\overline{\nu}_{\mu}$ and
$\nu_{\mu}$ events respectively by applying neutrino oscillations. At
the end of the events generation, we have events listed in full detail,
including the energy and direction $(E_\nu, \cos\theta_\nu, \phi_\nu)$
of the initial neutrino, the energy and direction of the final state
muon, $(E_\mu, \cos\theta_\mu, \phi_\mu)$ along with the sign of its
charge, and detailed information on all the final state hadrons produced
in the interaction. As detailed in Ref.~\cite{Mohan_2017}, we use the
information on the muon energy and direction $(E_\mu, \cos\theta_\mu)$,
and the total hadron energy, $E_{had}' \equiv E_\nu - E_\mu$, for the
analysis, while retaining the information on the neutrino energy and
direction $(E_\nu,\cos\theta_\nu)$, for later use in order to generate
the neutrino oscillation probabilities.

\subsection{Inclusion of detector response}

The events generated by the NUANCE neutrino generator give the true
values of the various parameters. However, in the actual detector, these
will be smeared depending on the detector resolutions. We therefore
smear the events according to the detector response studied in the
previous section. That is, we incorporate the efficiencies as well as
the resolutions in both energy and direction. In particular, we use the
look-up tables generated as described in the previous section to smear
the values of the muon energy and direction, as well as the hadron
energy, $(E_\mu, \cos\theta_\mu, E_{had}')$. The binning of the events
is done after including reconstruction efficiency of the muon and charge
identification efficiency of the detector. The {\em observed} muon CC
events are then given as
\begin{eqnarray} 
N^{tot}_{\mu^-}(E^{obs}_\mu,\cos\theta^{obs}_\mu,E_{had}^{'obs}) & = &
\epsilon_{rec} \epsilon_{cid} N_{\mu^-} +
\epsilon_{rec} (1-\epsilon_{cid}) N_{\mu^+}~, \\ \nonumber 
N^{tot}_{\mu^+}(E^{obs}_\mu,\cos\theta^{obs}_\mu,E_{had}^{'obs}) & = &
\epsilon_{rec} \epsilon_{cid} N_{\mu^+} + 
\epsilon_{rec} (1-\epsilon_{cid}) N_{\mu^-}~.
\end{eqnarray}
Here $\epsilon_{rec}$ and $\epsilon_{cid}$ are the reconstruction
efficiency and charge identification efficiency of the detector for
both $\mu^+$ and $\mu^-$; N$^{tot}_{\mu^-}$ and (N$^{tot}_{\mu^+})$
are the total oscillated CC $\nu_\mu$ ($\overline{\nu}_{\mu}$) events that will
be observed in a given $(E^{obs}_{\mu},\cos\theta^{obs}_{\mu},
E_{had}^{'obs})$ bin. Notice that since $\epsilon_{cid} < 1$, a few
$\mu^+$ events are mid-identified as $\mu^-$ events and vice versa.
However, $\epsilon_{cid} > 98$\% for $E_\mu \gtrsim 2$ GeV; hence this
contamination is small.

The ``theory" events were smeared as per the resolutions
corresponding to the incorrect magnetic field map by assuming the field
to be $B_{reco} = fB_{map}$, where $f = 0.95$, for instance. That is,
the mean and $\sigma$ of the muon energy are calculated based on the
value of $f$, as explained above. More details on the nature of the
smearing are given below.

The events are oscillated using the oscillation parameters given in
Table~\ref{tab:osc}. The normal ordering is assumed throughout
unless otherwise specified. The data was scaled to 10 years so all
results correspond to 10 years exposure at ICAL.

\subsection{Analysis}

The main goal of this study is to understand the impact of errors in
the magnetic field measurement in ICAL on the sensitivity to the
neutrino oscillation parameters, especially in the 2--3 sector that
atmospheric neutrinos are dependent on. We study the impact of two
different kinds of errors. In both cases, events are generated according
to the true magnetic field map as shown in Fig.~\ref{fig:field_map}. Hence
the ``data" that is to be fitted to ``theory" is always generated
according to the true map, as would be the case if there was real data
from ICAL. For the ``theory" events, there are two possibilities. One is
that there was a calibration error so that the measured magnetic field
is systematically higher or lower than the true one (which is represented
by the original magnetic field map in this analysis). Also, it is possible
that the local magnetic field where an event is generated is different
from the simulated value due to fluctuations in iron quality, composition
or deviations in the physical construction. The latter case is idealised
by assuming that the local magnetic field fluctuates randomly around the
``true" or map value. Hence we consider the following scenarios.

\begin{enumerate}

\item Considering the magnetic field map using the ``true" $B$-field
where the magnetic field that is generated using MAGNET software (see
Fig.~\ref{fig:field_map}) is used to reconstruct the events. The results
corresponding to use of the true map for both data generation and theory
analysis are already available \cite{ICAL:2015stm}.

\item Considering the impact of a systematic change in the $B$-field map
($f = $ constant). Here the magnetic field is systematically increased
or decreased constantly by a factor $f$ for the whole map and the
``theory" events are reconstructed according to this modified map.

\item Considering random Gaussian variation in the $B$-field map. Here the
local magnetic field where the event is generated via a random gaussian
which is centred around the true value as given by the field map, with a
width $\sigma = 5$\% about the central value. In this case the magnetic
field value is ramdomly varied by generating a random number with the
sigma 0.05 of the central value (true field value) of the field and the
``theory" events are reconstructed using the corresponding generated
field value, $fB$. The importance and usefulness of the linear
parametrisation developed in the previous section is now obvious: the
random gaussian number generates an arbitrary value of the scale factor
$f$ and the reconstructed muon energy is calculated based on linear
interpolation.

\end{enumerate}

Of course, it is possible that the realistic case will correspond to
inclusion of both systematic and measurement errors but we only consider
them individually in this analysis.

\subsection{$\chi^2$ analysis}

Both the ``data" and ``theory" events are scaled to 10 years (unless
otherwise specified) and binned in bins of observed muon energy and
direction $(E_\mu^{obs}, \cos\theta_\mu^{obs})$ and observed hadron
energy, $E_{had}^{'obs}$. The number and size of the bins were optimised
in the earlier study \cite{Mohan_2017} and used as-is in this
analysis.

Apart from the specific magnetic field variation
of interest, we consider five types of systematic errors which are given
in Table \ref{tab:pulls} and which are included using the method of pulls.
The flux normalization pull is the uncertainty in the assumed
(theoretical) energy dependence of the atmospheric neutrino flux and it
is calculated for each bin as follows:
\begin{equation}
\Phi_{\delta}(E) = \Phi_{0}(E)\left(\frac{E}{E_{0}}\right)^\delta
                 \simeq \Phi_{0}(E)\left(1 + \delta \ln
		 \frac{E}{E_{0}}\right)~.
\end{equation}
The uncertainty in angular flux is taken to be 5\% cos$\theta$,
that in the overall flux normalization is considered as 20\%, the
uncertainty in the cross section is considered to be 10\% and an overall uncertainty in detector response is taken to be 5\%.

\begin{table}[htp]
\centering 
\begin{tabular}{|c|c|} \hline  
Systematic error     & Error    \\ \hline \hline 
 Flux normalization  &  20\%    \\ \hline
 Shape uncertainty or tilt  & 5\%       \\\hline
 Zenith angle uncertainty   & 5\%        \\\hline
 Cross section uncertainty  & 10\%       \\\hline
 Detector response          & 5\%        \\\hline
\end{tabular}
\caption{The five types of systematic uncertainties included in the analysis.}
\label{tab:pulls}
\end{table}

The analysis closely follows that used in Ref.~\cite{Mohan_2017} and
the results of that work will be used as a baseline to understand the
effects of errors in the measurement of the magnetic field.

The loss of sensitivity due to using the incorrect field map during
reconstruction of the ``theory" events is determined through $\chi^2$
which is the sum of contributions from $\mu^+$ and $\mu^-$ observed
events:
\begin{eqnarray} 
\chi^2 & = & {\stackrel[\displaystyle {\xi_l^{\pm}}]{}{\hbox{min}}}
\sum^{N_{E^{obs}_{\mu}}}_{i=1}\sum^{N_{\cos\theta^{obs}_{\mu}}}_{j=1}
\sum^{N_{E'^{obs}_{had}}}_{k=1} \left\{ 2\left[ \left(T^{+}_{ijk}
- D^{+}_{ijk} \right) - D^{+}_{ijk} \ln \left( 
\frac{T^{+}_{ijk}}{D^{+}_{ijk}} \right) \right] + \right. \nonumber \\
& & \left. 2\left[\left(T^{-}_{ijk} - D^{-}_{ijk}\right) - D^{-}_{ijk}
\ln\left(\frac{T^{-}_{ijk}}{D^{-}_{ijk}}\right)\right] \right\} + 
\sum^{5}_{l^{+}=1} \xi^{2}_{l^{+}} + \sum^{5}_{l^{-}=1}
\xi^{2}_{l^{-}}~,
\label{chisq}
\end{eqnarray}
where T, D correspond to ``theory" and ``data", with the former
including the systematic errors through
\begin{eqnarray}
T^{+}_{ijk} & = & T^{0+}_{ijk}\left(1+\sum^{5}_{l^{+}=1} 
\pi^{l^{+}}_{ijk}\xi_{l^{+}}\right)~, \\ \nonumber
T^{-}_{ijk} &= & T^{0-}_{ijk}\left(1+\sum^{5}_{l^{-}=1}
\pi^{l^{-}}_{ijk}\xi_{l^{-}}\right)~,
\label{td-pi6xi6}
\end{eqnarray}
where the superscript '0' indicates the events from $\mu^\pm$ in the
absence of systematic errors. When the ``theory" events are generated
using the modified magnetic field map, the quality of the fit degrades
as quantified by
\begin{equation}
\Delta\chi^2(\lambda) = \chi^2(\lambda)-\chi^2_0~,
\end{equation}
with $\chi^2_0$ being the minimum value of $\chi^2$ when the true
magnetic field map is used to generate both ``data" and ``theory"
events. With no statistical fluctuations, $\chi^2_0 = 0$. Here, $\lambda$
refers to the sensitivity to any of the oscillation parameters such as
$\lambda = \sin^2\theta_{23}$.

\section{Results}

The precision reach for a parameter is defined as
\begin{equation}
P^{n\sigma} (p) \equiv \frac{\Delta V^p_n}{2V^p_0}~,
\label{eq:prec}
\end{equation}
where $\Delta V_n^p$ is the allowed range of the values of the parameter
$p$ at $n\sigma$, when the remaining parameters are marginalised over
their $3\sigma$ ranges, and $V^p_0$ is its central value. The precision
reach for each choice of $B$ field also depends on the minimum $\chi^2$
for the best-fit value for that choice.

The analysis is first performed for the case when the fitted magnetic
field is systematically 5\% ($0.95B$, $1.05B$)and 2\% ($0.98B, 1.02B$)
smaller/larger than the true values. The resulting $\chi^2$ is always
compared to the case where the true magnetic field map is used for both
data generation as well as for generating the theoretical rates. Hence,
what is plotted in the subsequent figures is the relative increase in
$\chi^2$ when there is a mismatch between the two maps. 

The analysis is then performed for the case when the fitted magnetic
field is randomly chosen as gaussian distributed around the true map
with a gaussian width of $\sigma= 5$\%. Again the resulting sensitivity
is compared to the case when the true map is used for both data and
theory.

\subsection{Sensitivity to the mixing angle $\theta_{23}$}

The theory value of $\sin^2\theta_{23}$ is kept fixed at different values
and the $\chi^2$ marginalised over the $3\sigma$ range of $\Delta m^2$.

\begin{figure}[htp]
 \centering
\includegraphics[width=0.49\textwidth]{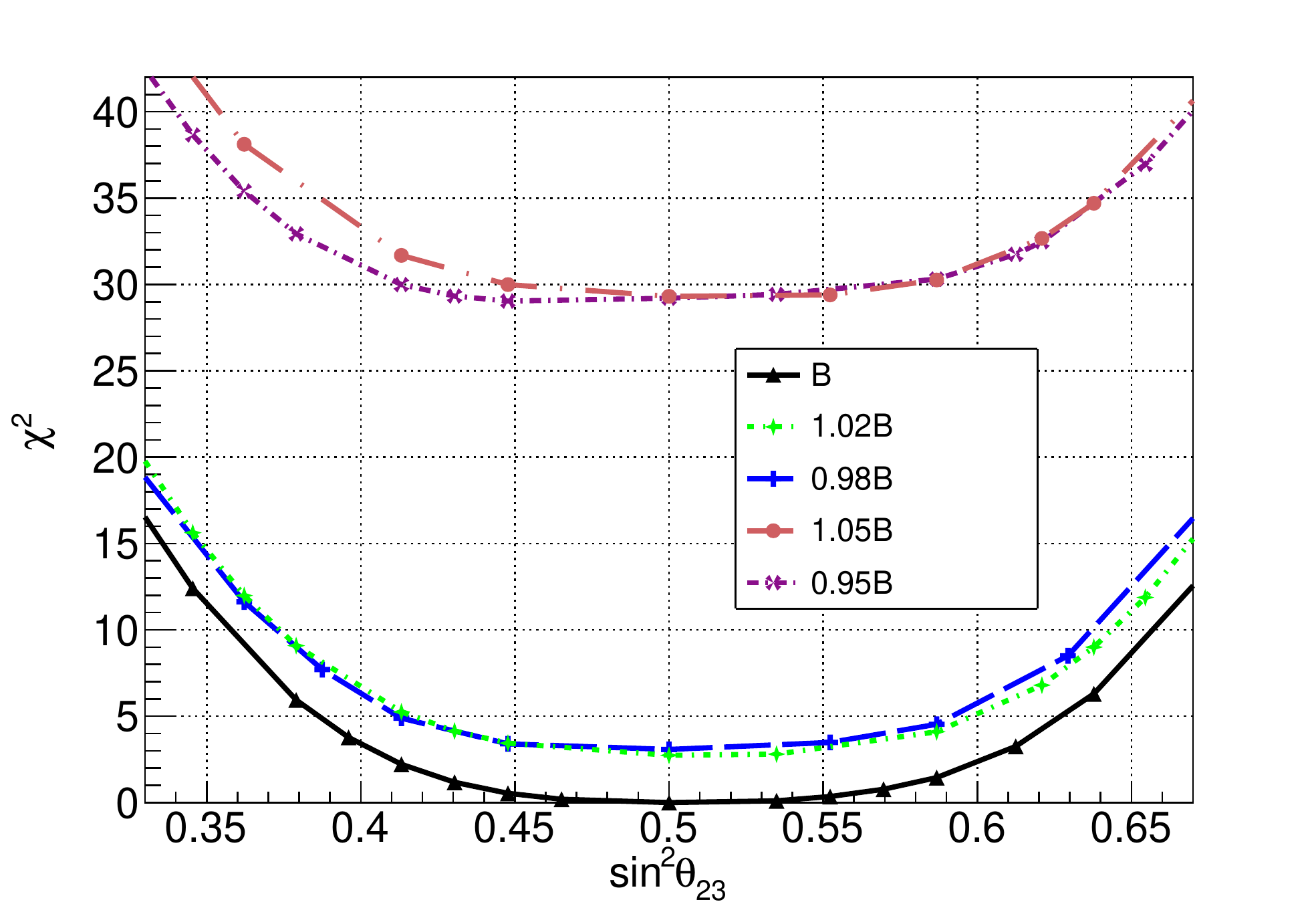}
\includegraphics[width=0.49\textwidth]{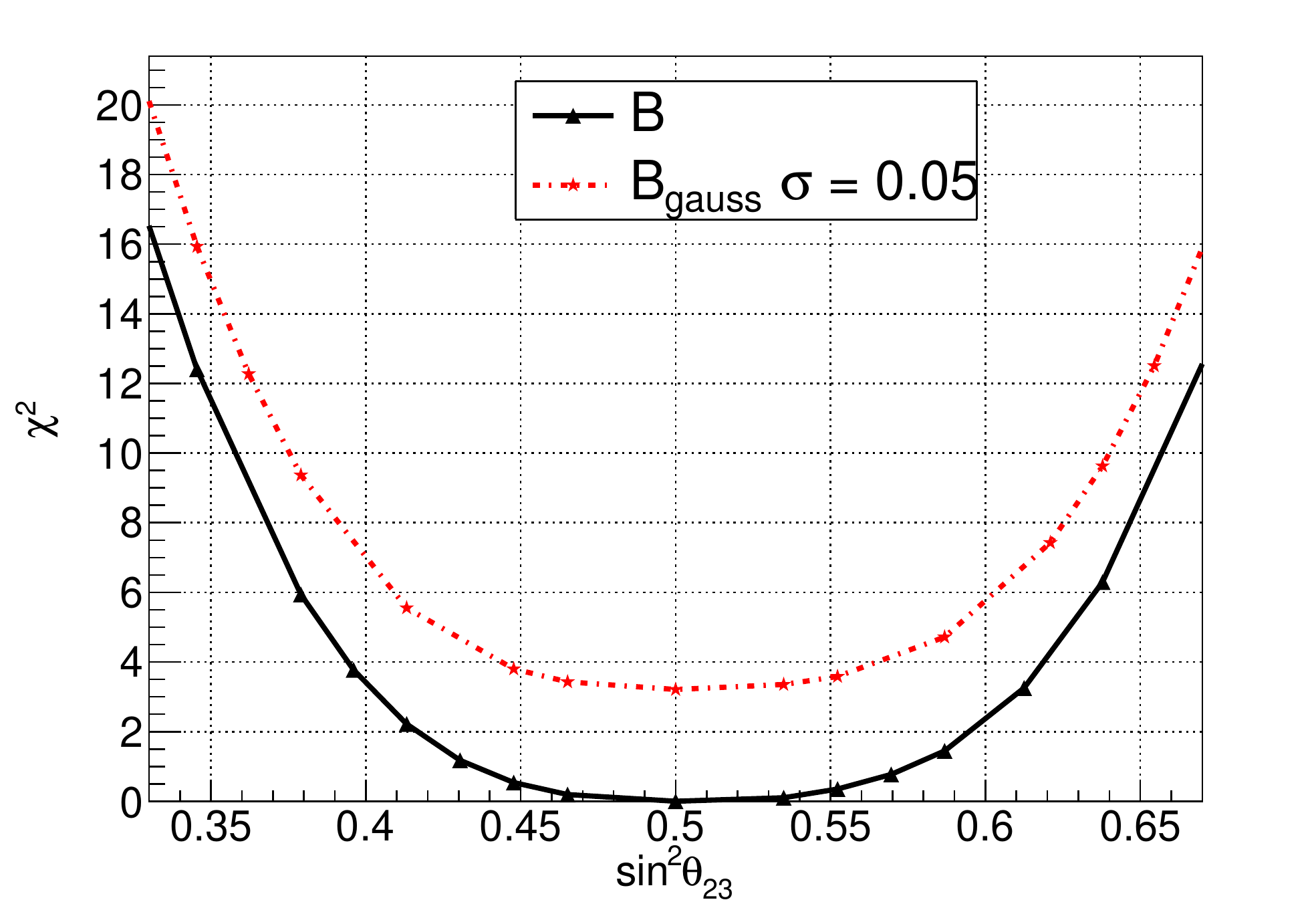}
\caption{Left: $\Delta \chi^2$ as a function of the theory value of
$\sin^2\theta_{23}$ when the true or data value is fixed to
$\sin^2\theta_{23}=0.5$; $\Delta m^2$ has been marginalised over its
$3\sigma$ range. The different curves correspond to theory events
generated with a systematically modified magnetic field map, modified
by a factor $f$, where $f = 1 \pm 2\%, 1\pm 5$\%. It can be seen that 5\% field
distortion drastically worsens the $\chi^2$ and hence the quality of
the fit. Right: The same plot for a gaussian variation of the magnetic
field map with width $\sigma=0.05$. In both plots the result with true
magnetic field map is shown as the solid black line. Note that the $y$
axis ranges are different for the two plots.}
\label{fig:t23}
\end{figure}

\paragraph{Systematic change in $B$ field}: From the left hand figure of Fig.~\ref{fig:t23}, it can be seen that for
a $\pm 2$\% systematic variation in the theory field map, there is some
worsening (by a few percent) of the precision measurement at $2\sigma$
($\Delta \chi^2 = 4$) of $\theta_{23}$ although the minimum $\chi^2$
is worse by about $\chi^{2,fB}_{min} - \chi^{2,B}_min \sim 3$, where
$\chi^{2,B}_{min}=0$ always by definition for the case when the magnetic
field maps match. For larger systematic errors of $\pm 5$\%, the
minimum $\chi^2$ drastically worsens; in addition, the best-fit value
moves away from the input data value. We will discuss these trends
later below when we discuss the simulataneous best-fit for
$\sin^2\theta_{23}$ and $\Delta m^2$. It be be noted here that poor
quality fits to the data may indicate potential errors in the
calibration of the magnetic field and this is an important insight when
the main ICAL comes online.

\paragraph{Gaussian variation in $B$ field}: For a 5\% Gaussian
variation around the true the sensitivity is reasonably similar to the
case when the true magnetic field is used for generating the theory,
as can be seen from the right side of Fig.~\ref{fig:t23}. In
particular, the best fit value remains the true value, which is to be
expected since the theory magnetic field map simply fluctuates around the
true value. Such small fluctuations may be caused by local variations
in the ICAL geometry, errors in cutting the iron plates, improper
alignment of the iron plates, etc. It appears that the tolerance for
such deviations is much better than when there are systematic errors in
the overall calibration of the magnetic field itself.

\subsection{Effect on sensitivity for $\Delta m^2$}
 
A similar study was carried out to determine the impact on the
sensitivity to the mass squared difference, $\Delta m^2$. Here the
$\chi^2$ was marginalised over the $3\sigma$ range of
$\sin^2\theta_{23}$ for various theory values of $\Delta m^2$.

\begin{figure}[htp]
 \centering
\includegraphics[width=0.49\textwidth,height=0.365\textwidth]{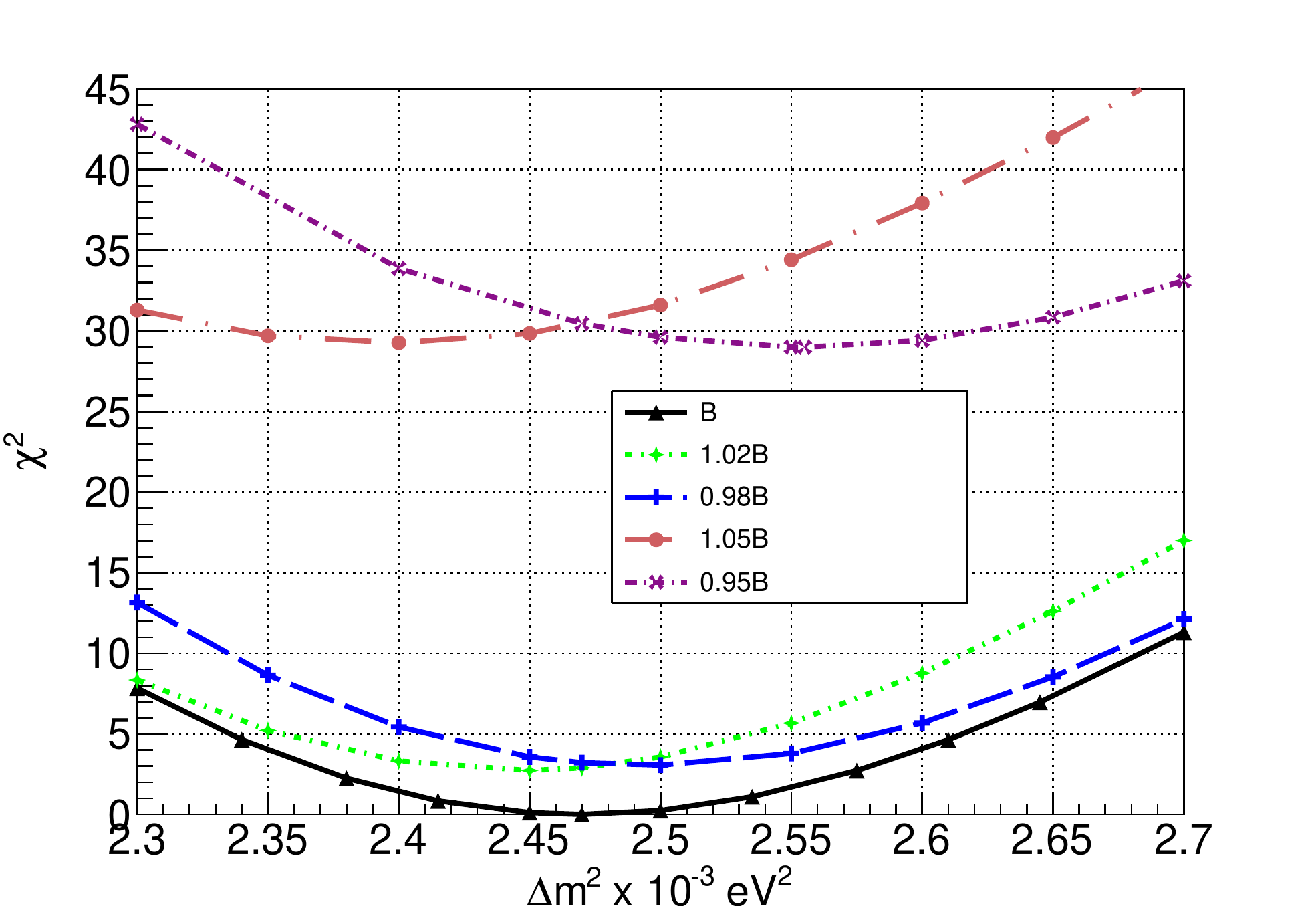}
\includegraphics[width=0.49\textwidth]{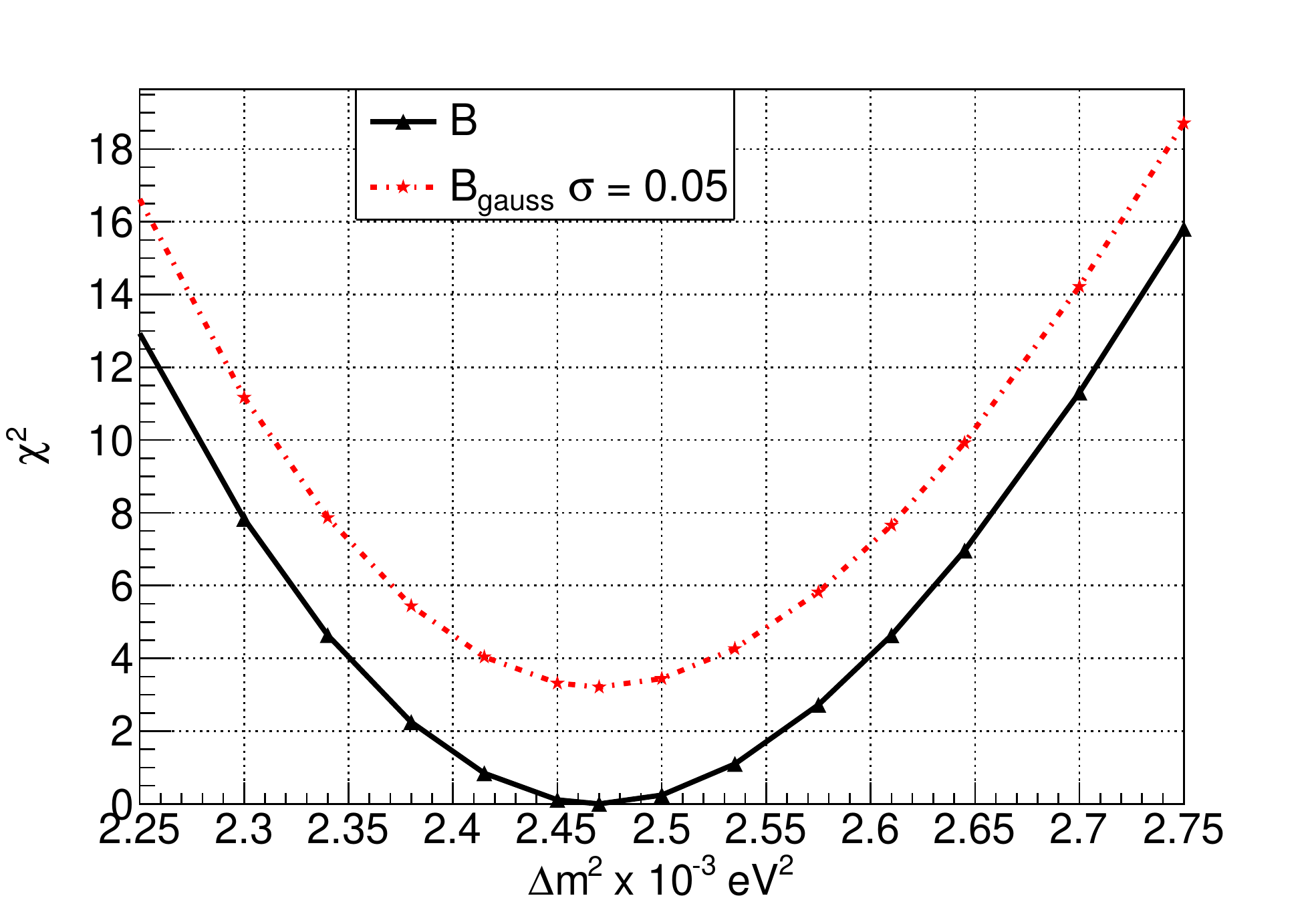}
\caption{Left: $\Delta \chi^2$ as a function of the theory value of
$\Delta m^2$ in units of eV$^2$ when the true or data value is fixed to
$\Delta m^2=2.47\times 10^{-3}$ eV$^2$; $\sin^2\theta_{23}$ has been marginalised over its $3\sigma$ range. The different curves correspond to theory events
generated with a systematically modified magnetic field map, modified
by a factor $f$, where $f = 1 \pm 2\%, 1\pm 5$\%. It can be seen that 5\% field
distortion drastically worsens the $\chi^2$ and hence the quality of
the fit. Right: The same plot for a gaussian variation of the magnetic
field map with width $\sigma=0.05$. In both plots the result with true
magnetic field map is shown as the solid black line. Note that the $y$
axis ranges are different for the two plots.}
\label{fig:dm2}
\end{figure}

\paragraph{Systematic change in $B$ field}: From the left hand figure
of Fig.~\ref{fig:dm2}, again, it can be seen that
$\pm 2$\% systematic changes in the field map did not cause
large changes in the sensitivity. Again, a $\pm 5$\% variation is beyond the
tolerance limit as the quality of fit worsens considerably. It is
interesting to note that unlike in the case of $\sin^2\theta_{23}$, the
deviation of the best-fit value from the true value is very visible
even for small deviations from the true field map. However, we will see
below that this is the case for $\sin^2\theta_{23}$ as well, although
not as clearly visible in Fig.~\ref{fig:t23}.

\paragraph{Gaussian variation in $B$ field}: The sensitivity does not
change significantly for a 5\% Gaussian variation of the field about
its true value, as can be seen from the right panel of
Fig.~\ref{fig:dm2}. Again, therefore, such fluctuations are more
tolerable for precision measurements of $\Delta m^2$ than calibration
errors in the magnetic field.

\subsection{Combined fit to $\sin^2\theta_{23}$ and $\Delta m^2$}

\begin{figure}[htp]
\centering
\includegraphics[width=0.6\textwidth]{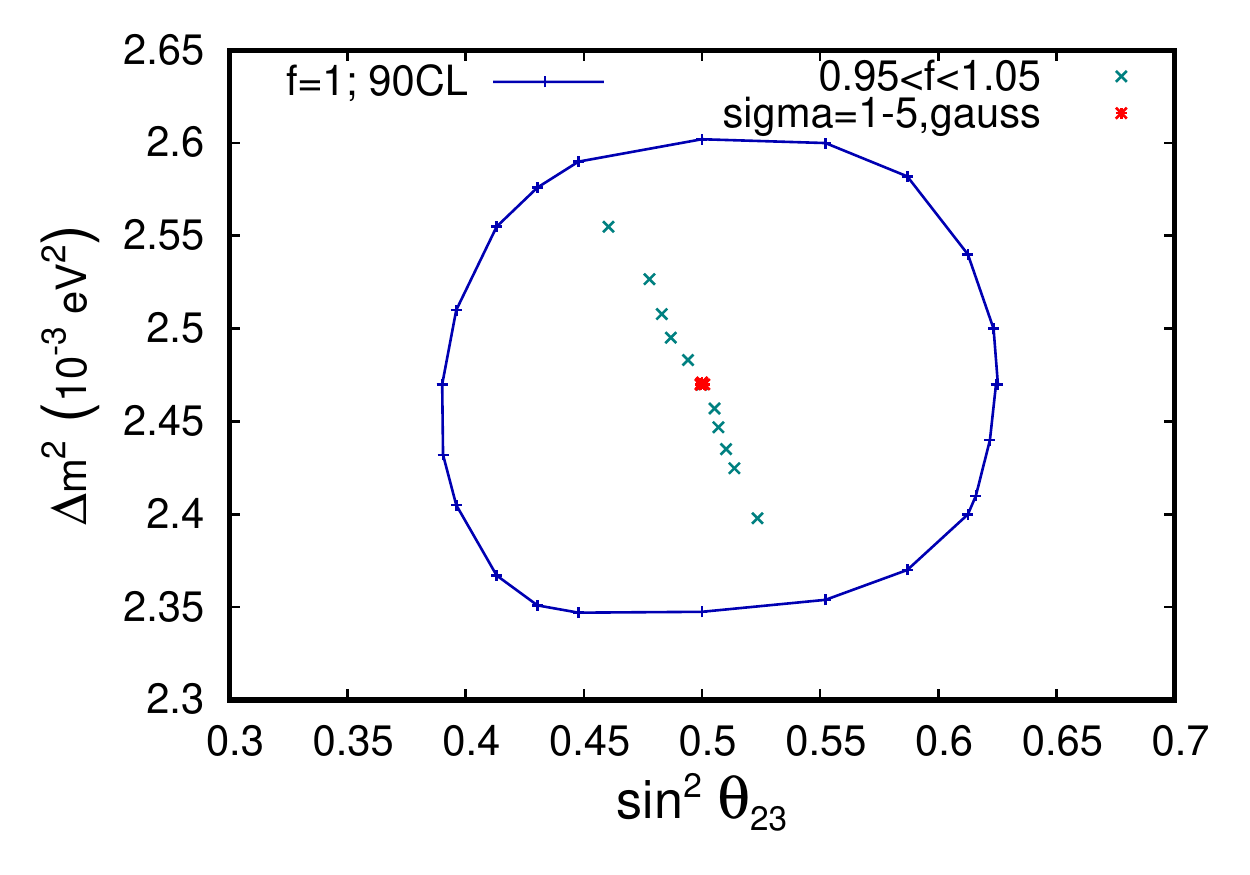}
\caption{Allowed 90\% CL two parameter contour in the
$\sin^2\theta_{23}$--$\Delta m^2$ plane when the true or data value
is fixed to $(\sin^2\theta_{23},\Delta m^2)=(0.5,2.47\times 10^{-3}$
eV$^2$) when the true magnetic field map is used to generate the theory
events. The green crosses from top left to bottom right correspond
to the best-fit values when the magnetic field map used in the theory
reconstruction is systematically different from that for the data by $f
= 0.95$--1.05 in steps of 1\% while the red square corresponds to the
best-fit values when a gaussian variation upto 5\% is used instead. See
text for details.}
\label{fig:central}
\end{figure}

In order to better understand the variation of the best fit values with
changing magnetic field maps, we have performed a simultaneous
2-parameter fit for each case. The results are shown in
Fig.~\ref{fig:central} for cases when there is a systematic change in
the magnetic field by a factor $\vert 1 - f\vert =2,5$\%, as well as
for the gaussian case. Shown in blue is the 90\% CL contour in the
$\sin^2\theta_{23}$--$\Delta m^2$ plane for the case when the true
field map is used to generate both the data and theory. This is the
ideal case\footnote{While the procedure followed is given in this
reference, the contour has been redrawn for the new central parameter
values that have been used in this work.} that has been discussed
earlier \cite{Mohan_2017}. The green crosses from top left to bottom
right mark the best fit values for $f = 0.95$--1.05 in steps of 1\%
respectively. The dense red square in the centre, at the position of
the true (data) values, corresponds to {\it all} the cases where the
magnetic field map is generated by a gaussian fluctuation of the field
around the central value, with width $\sigma=1$--5\% in steps of 1\%.

It can be seen that the gaussian fluctuation does not change the
best-fit value while systematic errors in the magnetic field map also
change both the $\sin^2\theta_{23}$ and $\Delta m^2$ values
systematically away from the true or input data values. The reason for
the trend in the best fit values for the case when there is a
systematic variation in the field is a consequence of the complicated
dependence of the oscillation probabilities on the two parameters of
interest. It directly arises from the fact that the reconstructed muon
momentum is systematically smaller/larger than the true value when the
modified field is systematically smaller/larger than the true map.
Since the flux falls severely with neutrino energy, the fact that the
cross section increases with energy is unable to compensate for this
loss; hence the muon events are peaked at low energies. When this data
is fitted using a larger/smaller field map, the peak in the events is shifted
to higher/smaller muon energies. The oscillation parameters required
to fit this distorted events distribution therefore change. The result
has a complicated dependence on the energy as well as direction due to
matter effects; see \cite{Indumathi:2006gr} for the dependence of both
$P_{\mu\mu}$ and $P_{e\mu}$ on the oscillation parameters as a function
of energy and direction of the incoming neutrino. The result however
is a straightforward trend in the best fit values as seen in the figure.

\begin{figure}[htp]
 \centering
\includegraphics[width=0.6\textwidth]{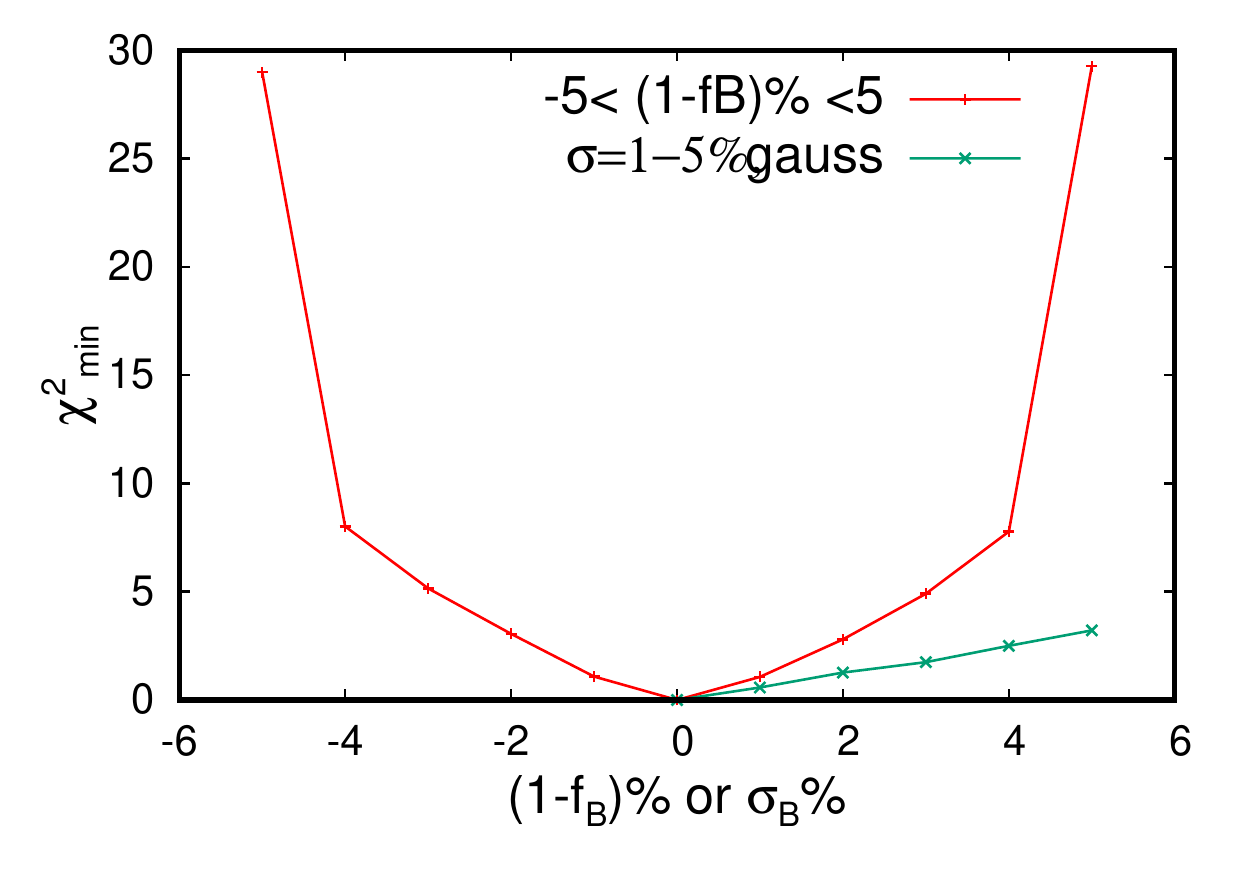}
\caption{The minimum $\chi^2$ value corresponding to the best-fit
values shown in Fig.~\ref{fig:central} for the simultaneous fit in the
$\sin^2\theta_{23}$--$\Delta m^2$ plane. The red curve with dot points
corresponds to the values when the magnetic field map
used in the theory reconstruction is systematically different by $\pm
5$\% in steps of 1\% while the green crosses correspond to the best-fit
values when a gaussian variation upto 5\% is used instead. See text
for details.}
\label{fig:chisq_central}
\end{figure}

Although the best-fit values correspond to increasingly poor $\chi^2$
values as the deviation from the true value increases it can be seen
from Fig.~\ref{fig:central} that variations up to $\pm 5$\% yield
best-fit values that still lie within the 90\% CL of the true result.
The minimum $\chi^2$ values are shown as a function of the deviations
from the true map in Fig.~\ref{fig:chisq_central}. It can be seen that
the quality of fit remains decent for gaussian fluctuations of the
field around the true value.

\subsection{Mass hierarchy sensitivity}

\begin{figure}[tbp]
\centering
\includegraphics[width=0.65\textwidth,height=0.5\textwidth]{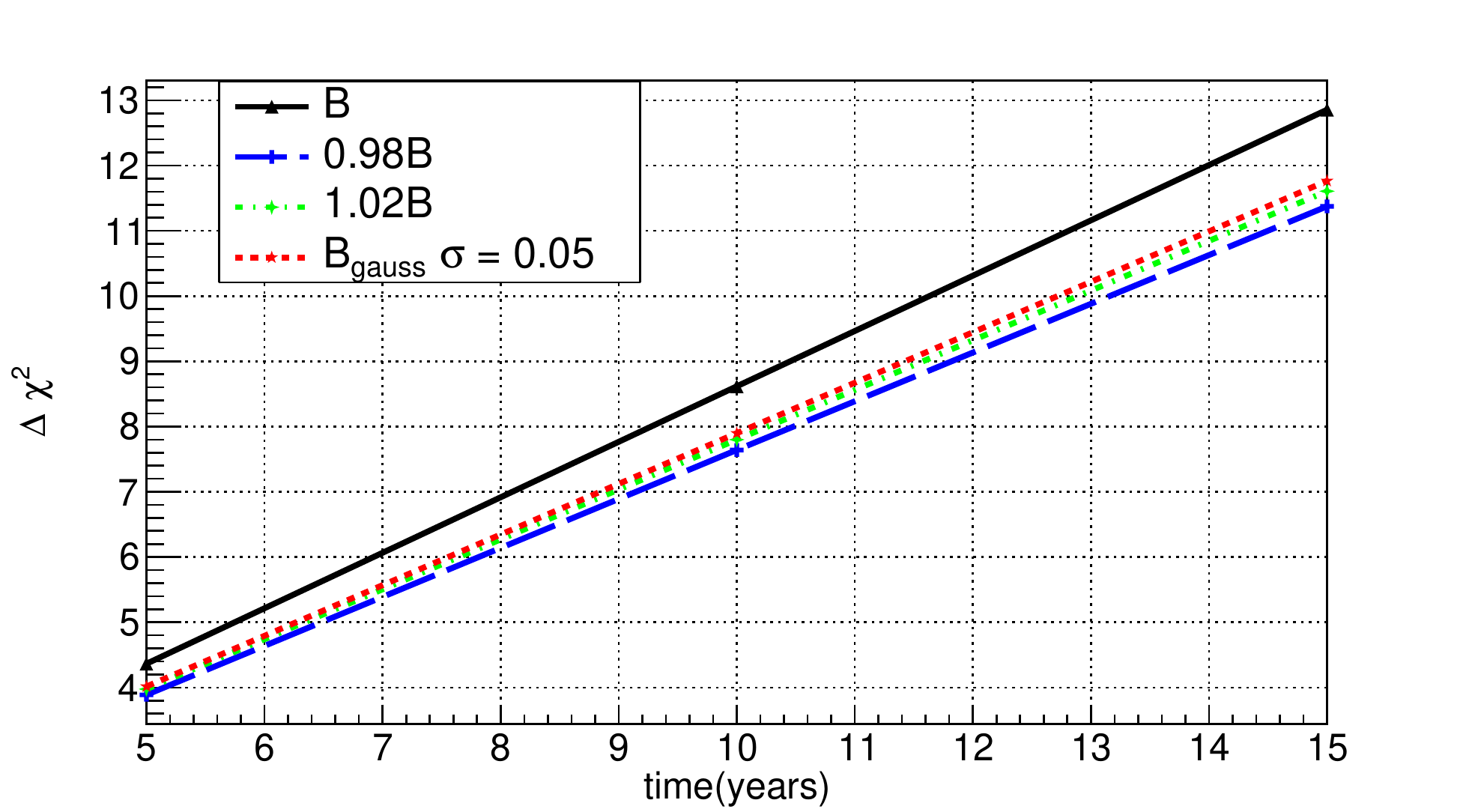}
\caption{Ability to discriminate against the wrong mass ordering as a
function of exposure time in years at ICAL with theory events being
generated according to true map (black solid line), showing a loss in
sensitivity when there is a $\pm 2$\%
systematic deviation in the map (green and blue dashed and dotted
lines), and a marginal improvement (red dotted line) when a 5\%
gaussian modified map is used.}
\label{fig:hier}
\end{figure}

Here events were generated using normal ordering (NO) and $\Delta
\chi^2$ was calculated by assuming the inverted ordering (IO) in the
theory. The results were marginalised over $3\sigma$ ranges of both
$\sin^2\theta_{23}$ and the magnitude of $\Delta m^2$. As expected,
the sensitivity of the ICAL to the mass ordering decreases with
smearing in the $B$-field. However, as can be seen from
Fig.~\ref{fig:hier}, the loss of sensitivity is marginal for a $\pm
2$\% systematic variation or a 5\% gaussian variation of the field,
with a $3\sigma$ hierarchy sensitivity being achievable in $11+$ years
of running of ICAL rather than $10+$ years with ideal field map.

\section{Discussion and conclusions}

We have presented, for the first time, a detailed simulations analysis of the
realistic case when the magnetic field map for ICAL is not precisely
known, and the consequent impact on the precision with which the
oscillation parameters, especially in the 2--3 sector can be
determined in this case. We have used the magnetic field measurements
made on the prototype mini-ICAL 85 ton detector which is an exact
scale model of the proposed ICAL detector to determine reasonable
ranges of errors, both in measurement and calibration, of the magnetic
field. First, we generated 250 sets of 10,000 muon events each at
different energies and angles in a Geant4 simulation of the ICAL
detector. The resulting muon tracks were then reconstructed and
analysed using (1) the true map, (2) maps that deviated from the true
map systematically by a constant factor $f$, being either larger than
or smaller than the true field. It was found that for small deviations
of the map, the deviations in the reconstructed muon momentum showed a
linear dependence on the deviation of the magnetic field, and was
non-linear for larger deviations. It was further found that these
variations were independent of the {\it region} in ICAL where the
event was generated. The muon momentum resolution also changed
linearly, although the variation of this parameter does not
significantly affect the results. It was also found that, within
reasonable errors, the reconstruction efficiency, charge-id
efficiency, and the angular resolution remained the same as with the
true magnetic field map.

This study was then used to parametrise the muon parameters (energy,
resolution functions, etc) in terms of the magnetic field parameter
$f$. This enabled analysis for arbitrary values of the magnetic
field. In addition, five systematic errors corresponding to pulls for
flux and cross section normalisation, flux angular distribution, flux
energy dependence or tilt, and detector parameters, were included in
the analysis. It was found that small systematic deviations from the
true field gave acceptable results with very little loss of
sensitivity to the 2--3 oscillation parameters $\sin^2\theta_{23}$ and
$\Delta m^2$ (or equivalently, $\Delta m^2_{32}$). However, large
variations of say 5\%led to very bad quality fits. It appears that
poor quality fits may be useful as an indicator of issues with
calibration of the magnetic field and is an important outcome of this
study for the main ICAL detector.

Random fluctuations of the magnetic field in different
regions of ICAL with gaussian widths of 5\% around the true magnetic
field value, on the other hand, gave results with very good quality
fits for both the 2--3 parameters; in addition the resulting best-fit
parameters were very close to the true input values, in contrast to
the case when there are systematic variations in the magnetic field.
Hence, errors, especially in the calibration of the magnetic field
map, may give rise to best-fit values which trend away from the true
values, depending on the size of the deviations.

As mentioned earlier, this is the first detailed study of the impact
of errors in the measurement of the magnetic field in ICAL on the
quality and correctness of the fits to the neutrino oscillation
parameters $\sin^2\theta_{23}$ and $\Delta m^2_{32}$. More data is
being currently taken at the prototype mini-ICAL detector in Madurai,
South India. In addition, detailed simulations studies of the magnetic
field map for both the main ICAL and mini-ICAL are on-going and will
be compared to the measured values. Both these will augment this
current first and preliminary analysis and allow for a more detailed
study of this crucial input to ICAL physics.

\paragraph{Acknowledgements}: We thank V.M. Datar and Amol Dighe for
a careful reading of the manuscript and many clarifications, and the
members of the ICAL collaboration for support and discussions.

\end{document}